\documentclass[journal]{IEEEtran}
\usepackage{ifpdf}
\usepackage{cite}
\usepackage{amsmath}
\usepackage{textcomp}
\usepackage{filecontents}
\usepackage{microtype}
\usepackage{float}
\usepackage{adjustbox}
\usepackage{booktabs,makecell,tabularx}
\usepackage{url}
\usepackage{booktabs}
\usepackage{subfigure}
\usepackage{algpseudocode}
\usepackage{algorithmicx}
\usepackage{array}
\usepackage{dblfloatfix}
\usepackage{url}
\usepackage[ruled,vlined]{algorithm2e}
\usepackage{amssymb,mathtools}

\begin{document}

\SetAlFnt{\small}
\SetAlCapFnt{\small}
\SetAlCapNameFnt{\small}

\title{\LARGE Low-Complexity Algorithm for Outage Optimal Resource Allocation in Energy Harvesting-Based UAV Identification Networks}
\author{Jae Cheol Park, 
        Kyu-Min Kang, 
        and Junil Choi,~\IEEEmembership{Senior Member,~IEEE}
\thanks{This work was supported by the ICT R\&D program of MSIT/IITP [2019-0-00499, Development of identification and frequency management technology of small drones at low altitute] and the National Research Foundation (NRF) Grant funded by the MSIT of the Korea Government (2019R1C1C1003638).  \it{(Corresponding author: Junil Choi.)}}
\thanks{J. C. Park is with the Radio \& Satellite Research Division, Electronics and Telecommunications Research Institute, Daejeon 34129, South Korea, and also with the School of Electrical Engineering, Korea Advanced Institute of Science and Technology, Daejeon 34141, South Korea (e-mail: jchpark@etri.re.kr).}
\thanks{K.-M. Kang is with the Radio \& Satellite Research Division, Electronics and Telecommunications Research Institute, Daejeon 34129, South Korea (e-mail: kmkang@etri.re.kr).}
\thanks{J. Choi is with the School of Electrical Engineering, Korea Advanced Institute of Science and Technology, Daejeon 34141, South Korea (e-mail: junil@kaist.ac.kr).}
}

\markboth{ }
{ }

\maketitle

\begin{abstract}
We study an unmanned aerial vehicle (UAV) identification network equipped with an energy harvesting (EH) technique. In the network, the UAVs harvest energy through radio frequency (RF) signals transmitted from ground control stations (GCSs) and then transmit their identification information to the ground receiver station (GRS). Specifically, we first derive a closed-form expression of the outage probability to evaluate the network performance. Then we obtain the closed-form expression of the optimal time allocation when the bandwidth is equally allocated to the UAVs. 
We also propose a fast-converging algorithm for time and the bandwidth allocation, which is necessary for the UAV environment with high mobility, to optimize the outage performance of EH-based UAV identification network. Simulation results show that the proposed algorithm outperforms the conventional bisection algorithm and achieves near-optimal performance.
\end{abstract}

\begin{IEEEkeywords}
Energy harvesting, outage probability, resource allocation, unmanned aerial vehicle (UAV), UAV identification.
\end{IEEEkeywords}

\IEEEpeerreviewmaketitle

\section{Introduction}
Unmanned aerial vehicles (UAVs) have recently attracted a great attention for various applications like aerial base station, aerial filming, and aerial delivery because of their mobility in the airspace.
Meanwhile, the UAVs have caused serious problems such as flying in restricted areas, and recording videos and taking pictures illegally. 
The United States and Europe announced regulations about UAV remote identification for public safety and security \cite{EASA,FAA}. 
According to Federal Aviation Administration (FAA) regulations,
a UAV in flight has to support a remote identification function that transmits certain identification and location information to the ground station periodically, 
and UAVs without the remote identification function can only fly within FAA-recognized identification area (FRIA) \cite{FAA}.
Many UAVs in operation, which do not have built-in remote identification function on them, will be equipped with a separate module for the remote identification function to comply with these regulations.
Since the separate module may have very limited battery capacity to make it light weighted, energy harvesting (EH) \cite{Liu20,Lu15} will be a promising technique for the UAV remote identification function.

Recently, the UAVs have been applied in EH networks to enhance performance \cite{Yang20,Park19,Xie19}.
In UAV-aided wireless powered communication networks (WPCN), 
both path planning and energy-minimization algorithms were proposed to minimize the total energy consumption of UAVs while accomplishing the minimal data transmission requests of users \cite{Yang20}. 
To maximize the uplink minimum throughput, 
the UAV trajectory design and the transmission resource allocation 
were jointly optimized under the UAV maximum speed constraint and the user's energy neutrality constraint \cite{Xie19}. 
The UAV trajectories, uplink transmit power, and time resource-allocation algorithms were jointly optimized to maximize the minimum throughput in UAV-aided WPCN \cite{Park19}.
The iterative user association, computation capacity, and location allocation algorithms were proposed to minimize the sum power consumption of a UAV-enabled mobile edge computing (MEC) network  \cite{Yang19}.
However, the performance of the UAV identification function has not been investigated yet.

\begin{figure} 
\centering
\includegraphics[width=6.2 cm]{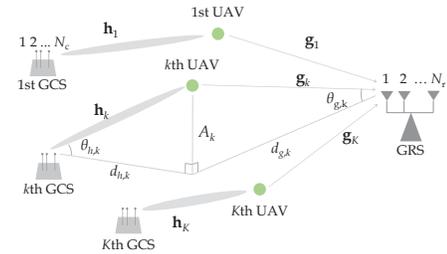} 
\caption{EH-UAV identification network.}
\vspace{-10pt}
\label{sys}
\end{figure}

In this letter, we investigate an EH-based UAV (EH-UAV) identification network 
in which the UAVs harvest energy through radio frequency (RF) signals transmitted from ground control stations (GCSs) and then transmit their identification information to the ground receiver station (GRS).  
The time and bandwidth resource allocation is considered to balance the harvested energy and the achievable rate of UAVs.
The contributions of this letter are summarized as follows.
\begin{itemize}
    \item We derive a closed-form expression of the outage probability and provide the optimal time-allocation factor for 
    equal-bandwidth allocation among UAVs.
    \item We propose a fast-converging algorithm, which is necessary for the UAV environment with high mobility, for time and bandwidth allocation to optimize the outage probability.
    \item Simulation results show that the proposed algorithm provides near-optimal performance with fewer iterations in 
    comparison to the conventional bisection algorithm.
\end{itemize}

\section{System Model}\label{sec2}
As shown in Fig. \ref{sys}, we consider an EH-UAV identification network
consisting of $K$ UAV-GCS pairs and one GRS.
The single-antenna UAV harvests energy through RF signals transmitted from its associated GCS with $N_\mathrm{c}$ antennas, 
and then transmits its identification information to the GRS with $N_\mathrm{r}$ antennas consuming the harvested energy. 

The channel from the $k$th GCS to the $k$th UAV is expressed as $\mathbf{h}_{k} =[h_{1,k},\cdots,h_{N_\mathrm{c}, k} ]^{\mathrm{T}} \in \mathbb{C}^{N_\mathrm{c} \times 1}$,
and each element is modeled as frequency flat and block Nakagami-$m$ fading with an integer parameter $m_{h,k}$ and an average parameter $\lambda_k = 10^{-\ell_{h,k}/10}$. 
Here, the air-to-ground (A2G) path-loss between the $k$th GCS and the $k$th UAV with an altitude $A_k$ is denoted by $\ell_{h,k}$, and it is given by \cite{Khuwaja18}
\begin{align} \label{pl}
    \ell_{h,k} =& \frac{\eta_{\text{LoS}} - 
    \eta_{\text{NLoS}}}{1+ a \exp\left[-b \left( \theta_{h,k} - a \right) \right]} 
    + 10\log\left(\sqrt{d_{h,k}^2+A_k^2}\right) \nonumber \\ 
    &+ 20\log\left( \frac{4\pi f_\mathrm{c}}{c} \right) + \eta_{\text{NLoS}},
\end{align}
where $a$ and $b$ are constants depending on environments such as suburban, urban, dense-urban, and high-rise urban. 
The mean additional losses for line-of-sight (LoS) and non-line-of-sight (NLoS) terms are expressed as $\eta_{\text{LoS}}$ and $\eta_{\text{NLoS}}$, respectively. 
The horizontal distance and elevation angle between the $k$th GCS and the $k$th UAV are $d_{h,k}$ and $\theta_{h,k}$, respectively, the carrier frequency is $f_\mathrm{c}$, and the speed of light is $c$. 

The channel from the $k$th UAV to the GRS is denoted as $\mathbf{g}_{k}  = [g_{1,k},\cdots,g_{N_\mathrm{r},k} ]^{\mathrm{T}} \in \mathbb{C}^{N_\mathrm{r} \times 1}$, and each element is also modeled as frequency flat and block Nakagami-$m$ fading with an integer parameter $m_{g,k}$ and an average parameter $\mu_k = 10^{-\ell_{g,k}/10}$.
The A2G path-loss $\ell_{g,k}$ between the $k$th UAV and the GRS can be calculated from (\ref{pl}) 
with the elevation angle $\theta_{g,k}$, the horizontal distance $d_{g,k}$, and the altitude $A_k$.
Note that the A2G path-losses $\ell_{h,k}$ and $\ell_{g,k}$ decrease as the altitude of UAVs up to a certain level and then increase.

\begin{figure} 
\centering
\includegraphics[width=6.0 cm]{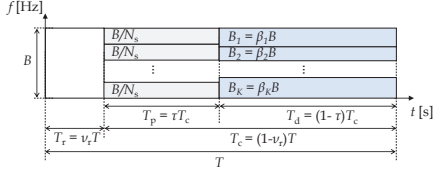} 
\caption{Resource block of the EH-UAV identification network.}
\vspace{-10pt}
\label{res}
\end{figure}

Fig. \ref{res} illustrates the resource block of the EH-UAV identification network.
The block time is given by $T=c/(\hat{V}f_\mathrm{c})$
where $\hat{V}=\max \{V_1, V_2, \cdots, V_K\}$ denotes the maximum velocity of UAVs, and $V_k$ means the velocity of the $k$th UAV. 
The block time is divided into three phases, namely, 
the resource-allocation phase (RAP) $T_\mathrm{r}=\nu_\mathrm{r} T$ with the normalized duration of the RAP $\nu_\mathrm{r} \in [0, 1)$, 
the power-transfer phase (PTP) $T_\mathrm{p}=\tau (1-\nu_\mathrm{r}) T$ with the time-allocation factor (TAF) $\tau \in (0,1)$, and the data-transmission phase (DTP) $T_\mathrm{d}=(1-\tau) (1-\nu_\mathrm{r}) T$.
The total bandwidth $B$ is divided into $N_\mathrm{s}$ sub-bands for the frequency hopping (FH) technique with the bandwidth $B/N_\mathrm{s}$ in the PTP 
and $K$ sub-bands for frequency division multiple access (FDMA) in the DTP. 
The bandwidth of the $k$th UAV for the DTP is expressed as $B_k=\beta_k B$, where $\beta_k \in (0,1) $ is the bandwidth-allocation factor (BAF) for the $k$th UAV with $\sum_{k=1}^{K}\beta_k=1$. 

In the RAP, 
the GRS computes the TAF and BAF based on a specific allocation algorithm where the duration of the RAP depends on the required iterations for the allocation algorithm to converge.
The normalized duration of the RAP is expressed as
$\nu_\mathrm{r} = T_\mathrm{r}/T $.

In the PTP, the GCS transmits control signals to the UAV, 
and the UAV exploits these signals to adjust the flight and to harvest energy. 
GCSs utilize the FH technique to avoid interference from other GCS's signals.
Note that the FH patterns of GCSs do not overlap,
and GCSs use a maximum ratio transmission (MRT) to extend the flight distance of UAVs.
We assume the GCS has perfect channel state information (CSI) of the $k$th GCS and UAV link. 
The received signal at the $k$th UAV is expressed as 
\begin{align}
r_{k} = \sqrt{p_{\mathrm{c},k}}\mathbf{h}_{k}^{\mathrm{H}}\mathbf{w}_k  x_{\mathrm{c},k}  + n_{k},  
\end{align}
where $p_{\mathrm{c},k}$ denotes the transmit power of the $k$th GCS, 
and $x_{\mathrm{c},k}$ denotes the control signal of the $k$th GCS with $\mathcal{E}[|x_{\mathrm{c},k} |^2]=1$, $\mathbf{w}_k =\mathbf{h}_k/||\mathbf{h}_k|| $ is the MRT beamformer at the $k$th GCS, and $n_k$ is the additive white Gaussian noise (AWGN) at the $k$th UAV with $\mathcal{E}[|n_k|^2]=\sigma^2$. 
It is assumed that the noise power is negligible compared to the control signal.
The harvested energy at the $k$th UAV is then expressed as 
\begin{align}
E_{k}  = \zeta p_{\mathrm{c},k} ||\mathbf{h}_{k} ||^2  \frac{B}{N_\mathrm{s}} T_\mathrm{p},
\end{align}
where $\zeta$ is the energy conversion efficiency \cite{Liu20,Lu15}.

In the DTP, the UAVs transmit their identification information to the GRS with the harvested energy. 
Each UAV utilizes the bandwidth $B_k$ for $k=1,...,K$ to avoid interference from other UAV's signals. 
The average transmit power of the $k$th UAV is expressed as
\begin{align}
p_{\mathrm{i},k}  = \frac{E_{k} }{B_k T_\mathrm{d}}  = \frac{\tau }{\beta_{k} (1-\tau) N_\mathrm{s}}
 \zeta p_{\mathrm{c},k} ||\mathbf{h}_{k} ||^2
\end{align}
with the BAF of the $k$th UAV $\beta_k$.
The received signal at the GRS on $B_k$ is expressed as
\begin{align}
\mathbf{y}_{k}  = \sqrt{p_{\mathrm{i},k} } \mathbf{g}_{k} x_{\mathrm{i},k}  + \mathbf{n}_{k},
\end{align}
where $x_{\mathrm{i},k} $ is the identification signal of the $k$th UAV with $\mathcal{E}[|x_{\mathrm{i},k} |^2]=1$,
and $\mathbf{n}_{k}=[n_{1,k},\cdots,n_{N_\mathrm{r},k}]^{\mathrm{T}} \in \mathbb{C}^{N_\mathrm{r} \times 1}$ is the AWGN with $\mathcal{E}[\mathbf{n}\mathbf{n}^{\mathrm{H}}] =\sigma^2\mathbf{I}_{N_{\mathrm{r}}}$. 
To maximize the instantaneous signal-to-noise ratio (SNR), the GRS performs a maximum ratio combining (MRC) to the received signal with
$\mathbf{q}_k ={\mathbf{g}_{k} }/{||\mathbf{g}_{k} ||}$. 
After the MRC, the combined signal at the GRS is given by 
\begin{align}
z_{k}  &= \mathbf{q}_k^{\mathrm{H}} \mathbf{y}_{k} = \sqrt{p_{k} } ||\mathbf{g}_{k} ||x_{\mathrm{i},k}  + \frac{\mathbf{g}_{k}^{\mathrm{H}} }{||\mathbf{g}_{k} ||}\mathbf{n}_{k}.
\end{align}
The SNR of the $k$th UAV at the GRS is written as
\begin{align}
 \frac{p_{k} }{\sigma^2}||\mathbf{g}_{k} ||^2=  \frac{\tau}{\beta_k(1-\tau)}{\gamma}_k, 
\end{align}
where ${\gamma}_k =\rho_k ||\mathbf{g}_{k} ||^2 ||\mathbf{h}_{k} ||^2$  with $\rho_k=\zeta \frac{p_{\mathrm{c},k}}{N_\mathrm{s} \sigma^2}$. 
The rate of the $k$th UAV is obtained as
\begin{align}
R_{k}(\beta_k, \tau)  =
 \beta_k (1-\tau)\nu_\mathrm{c} \log_2\left(1+\frac{\tau}{\beta_k(1-\tau)}{\gamma}_k \right), 
\end{align}
where $\nu_\mathrm{c} = 1-\nu_\mathrm{r}$ denotes the normalized duration for the PTP and the DTP.

\section{Outage Probability}
In this section, we derive the closed-form outage probability of the EH-UAV identification network.
The rate of the $k$th UAV should be higher than the required rate $R_\mathrm{a}$ to transmit the $k$th UAV's identification information to the GRS successfully. 
We first define the outage of the EH-UAV identification network when one of UAVs cannot support the required rate as 
\begin{align} \label{Po}
\mathcal{P}_o(\boldsymbol{\beta}, \tau, R_\mathrm{a}) = \Pr \left[\check{R}(\boldsymbol{\beta}, \tau)<R_\mathrm{a} \right] = F_{\check{R}(\boldsymbol{\beta}, \tau)} (R_\mathrm{a}),
\end{align}
where $\check{R}(\boldsymbol{\beta}, \tau)=\min \{R_1(\beta_1,\tau),~\cdots,~ R_K(\beta_K, \tau) \}$
is the minimum rate of UAVs with the BAF $\boldsymbol{\beta}=\left[\beta_1,\cdots, \beta_K\right]$ and the TAF $\tau$.

As in \cite{order}, the outage probability can be formulated as 
\begin{align} \label{F_check_r}
F_{\check{R} (\boldsymbol{\beta}, \tau) }(R_\mathrm{a})= 
1- \prod_{k=1}^{K}\left[1-F_{R_{k}(\beta_k, \tau)}(R_\mathrm{a})\right],
\end{align}
where $F_{R_k (\beta_k, \tau)}(\cdot)$ denotes the cumulative distribution function (CDF) of $R_k (\beta_k, \tau)$. 
It can be reformulated as
\begin{align} \label{Fr}
F_{R_k (\beta_k, \tau)}(R_\mathrm{a})&= \Pr \left[ R_k (\beta_k, \tau) <R_\mathrm{a} \right] \nonumber \\
&=F_{\gamma_k}\left(\frac{\beta_k(1-\tau)}{\tau} \left(2^{\frac{R_\mathrm{a}}{\beta_k(1-\tau)\nu_\mathrm{c}}} -1 \right ) \right), 
\end{align}
where $F_{\gamma_k}(\cdot)$ is the CDF of $\gamma_k$. Note that $\gamma_k$ is the product of two independent 
non-identical gamma random variables. 
As in \cite{Plazaola20}, the CDF of $\gamma_k$ is obtained by
\begin{align} \label{F_gamma}
    &F_{\gamma_k}(x)= 1 - \sum_{m=0}^{m_{h,k}N_\mathrm{c}-1} 
    \frac{2}{m! \Gamma(m_{g,k} N_\mathrm{r})} \nonumber \\ 
    &\times\left( \frac{x}{ \rho_k\lambda_k \mu_k} \right)^{\frac{m+m_{g,k}N_\mathrm{r}}{2}} \mathcal{K}_{m_{g,k} N_\mathrm{r}-m} \left( \sqrt{ \frac{4x}{ \rho_k\lambda_k \mu_k} } \right),
\end{align}
where $\Gamma(\cdot)$ is the gamma function and $\mathcal{K}_\alpha(\cdot)$ is the modified Bessel function of the second kind with an integer $\alpha$.
Finally, the closed-form expression of the outage probability for the EH-UAV identification network is derived as 
\begin{align} \label{Fana}
   &\mathcal{P}_o(\boldsymbol{\beta}, \tau, R_\mathrm{a}) = 1 - \prod_{k=1}^{K}\Bigg[ \sum_{m=0}^{m_{h,k}N_\mathrm{c}-1} 
    \frac{2}{m! \Gamma(m_{g,k} N_\mathrm{r})} \nonumber  \\
     &\times \left( \frac{ \beta_k (1-\tau)\left(2^{\frac{R_\mathrm{a}}{\beta_k (1-\tau)\nu_\mathrm{c}}}-1\right)}{ \tau\rho_k\lambda_k \mu_k} \right)^{\frac{m+m_{g,k}N_\mathrm{r}}{2}}  \nonumber \\
     &\times \mathcal{K}_{m_{g,k} N_\mathrm{r} -m} \left( \sqrt{ \frac{4\beta_k (1-\tau)\left(2^{\frac{R_\mathrm{a}}{\beta_k (1-\tau)\nu_\mathrm{c}}}-1\right)}{ \tau\rho_k\lambda_k \mu_k} } \right) \Bigg].
\end{align}

From the result of (\ref{Fana}), it is clear that the outage probability is monotonically increasing with
\begin{align}
    X_k(\beta_k, \tau) = \frac{\beta_k(1-\tau)}{\tau } \left(2^{\frac{R_\mathrm{a}}{\beta_k(1-\tau)\nu_\mathrm{c}}} -1 \right ).
\end{align}
Note that $X_k(\beta_k, \tau)$ is a concave function of $\tau$ with a given $\beta_k$ 
and a decreasing function of $\beta_k$ with a given $\tau$. 
Therefore, the optimal TAF $\tau^{*}$ can be obtained by solving $\frac{\partial }{\partial \tau}X_k(\beta_k,\tau)=0$
with the given BAF $\boldsymbol{\beta}$. 
As a special case, the optimal TAF for the equal-bandwidth allocation among UAVs with the BAF $\boldsymbol{\beta}=[1/K, \cdots, 1/K]$ is expressed as  
\begin{align} \label{t_s}
    \tau^{*} = 1- \frac{KR_\mathrm{a} \ln2}{1+ KR_\mathrm{a} \ln2 +\mathcal{L}\left(-e^{-1} 2^{-K R_\mathrm{a}} \right) },
\end{align}
where $\mathcal{L}(\cdot)$ denotes the Lambert-W function.

\section{Outage Optimal Resource Allocation}
In this section, we present the design of the resource allocation to optimize the outage probability of the EH-UAV identification network
and propose the fast-converging algorithm for resource allocation.

\subsection{Optimal Resource Allocation}
According to (\ref{Po}), the outage probability of the EH-UAV identification network depends on the minimum rate of UAVs. 
To optimize the outage probability, 
the GRS performs resource allocation to maximize the minimum rate of UAVs. The optimal outage probability is defined as 
\begin{align}
\mathcal{P}_o^{\star} (\boldsymbol{\beta}^\star, \tau^\star, R_\mathrm{a}) 
= \Pr[\check{R}(\boldsymbol{\beta}^\star, \tau^\star)<R_\mathrm{a}],
\end{align}
where 
$\check{R}(\boldsymbol{\beta}^\star, \tau^\star)=\min \{R_1(\beta^{\star}_1,\tau^{\star}),~\cdots,~ R_K(\beta^{\star}_K,\tau^{\star}) \}
$ is the maximized minimum rate of UAVs with the optimal BAF $\boldsymbol{\beta}^{\star}$ and the optimal TAF $\tau^{\star}$. 
The maximized minimum rate of UAVs can be obtained by solving the following optimization problem:
\begin{align} \label{op1}
&\max_{\boldsymbol{\beta}, \tau} \left\{  \min \left\{ R_{1} ({\beta_1}, \tau), \cdots, R_{K} ({\beta_K}, \tau) \right\} \right\}  \\
&\text{subject to}~ 0<\tau<1, \nonumber \\
&~~~~~~~~~~~~ \sum_{k=1}^{K}\beta_k = 1, \nonumber \\
&~~~~~~~~~~~~ 0<\beta_k <1,~ k=1,...,K. \nonumber 
\end{align}
The global optimal solution can be obtained by an exhaustive search method with excessive complexity, which is not practical due to the short length block time of UAV environments. 
Note that the duration of RAP determines the outage performance in the short block time condition. 
Therefore, 
it is important to consider the computational complexity of an resource-allocation algorithm 
to optimize the outage performance in UAV flight situations.

\subsection{Proposed Resource Allocation}
We designed the resource allocation with low-complexity to reduce the duration of the RAP.
The proposed algorithm is composed of two phases. 
In the first phase, the optimal TAF of the proposed algorithm can be obtained as
\begin{align} \label{op2-1}
\tau^o = \arg\max_{0<\tau<1} \left\{ \min \left\{ R_{1} ({\beta_1}, \tau), \cdots, R_{K} ({\beta_K}, \tau) \right\}  \right\}.
\end{align}
The objective function of (\ref{op2-1}) is the minimum rate of UAVs, which can be written as
\begin{align} \label{Rmin}
\check{R}(\boldsymbol{\beta}, \tau) = \beta_{\check{k}} (1-\tau)\nu_\mathrm{c}\log_2\left(1+\frac{\tau}{\beta_{\check{k}}(1-\tau)}{\gamma}_{\check{k}} \right),
\end{align}
where $\check{k}$ is the index of the minimum rate UAV. 
Because the minimum rate of UAVs satisfies $\check{R}(\boldsymbol{\beta}, \tau) \leq R_k(\beta_k, \tau)$ for $k=1,...,~K$ and $\tau \in (0,1)$,  
(\ref{op2-1}) is equivalent to
\begin{align}\label{op2-2}
\tau^o = \arg \max_{0 < \tau < 1} \left\{ \check{R}(\boldsymbol{\beta}, \tau) \right\}.
\end{align}
The minimum rate of UAVs $\check{R}(\boldsymbol{\beta}, \tau)$
is the concave function of the TAF $\tau$ with the given BAF $\boldsymbol{\beta}$. 
Hence, the optimal TAF $\tau^o$ always exists in $0<\tau<1$. 
Although (\ref{op2-2}) can be directly solved using the conventional bisection algorithm, 
it is possible to develop a low-complexity iterative time-allocation algorithm using the structure of (\ref{Rmin}). 

In the $n$th iteration of the first phase, the TAF is given by  
\begin{align}
    \tau_n = \frac{\tau_\mathrm{l} + \tau_\mathrm{u}}{2},
\end{align}
where $\tau_\mathrm{l}$ and $\tau_\mathrm{u}$ denote the lower and upper bounds of TAF $\tau$, respectively. 
For the next iteration, the lower bound of TAF $\tau_\mathrm{l}$ is updated to $\tau_n$ when
\begin{align} \label{thr_bisec}
 \frac{\partial \check{R}(\boldsymbol{\beta}, \tau)}{\partial \tau}\bigg|_{\tau=\tau_n} &=  - \beta_{\check{k}}\nu_\mathrm{c} \log_2 \left(1+ \frac{\tau_n}{\beta_{\check{k}}(1-\tau_n)}\gamma_{\check{k}} \right) \nonumber \\
& + \frac{\nu_\mathrm{c}\beta_{\check{k}}\gamma_{\check{k}}}{\ln 2 \left( \beta_{\check{k}}(1-\tau_n) + \tau_n \gamma_{\check{k}} \right)}
\end{align}
is positive and the upper bound of TAF $\tau_\mathrm{u}$ is updated to $\tau_n$  
when (\ref{thr_bisec}) is negative.
The time-allocation algorithm will be performed repeatedly until $\tau_\mathrm{u} - \tau_\mathrm{l} \leq \epsilon$ with a predefined threshold $\epsilon$.

In the second phase, the optimal BAF of the proposed algorithm can be obtained 
\begin{align} \label{op4}
&\boldsymbol{\beta}^o=\arg\max_{\boldsymbol{\beta}}\left\{  \min \left\{ R_{1} ({\beta_1}, \tau^o), \cdots, R_{K} ({\beta_K}, \tau^o) \right\} \right \}   \\
&\text{subject to}~ \sum_{k=1}^{K}\beta_k = 1,  \nonumber  \\
&~~~~~~~~~~~~~ 0<\beta_k <1,~ k=1,...,K. \nonumber 
\end{align}
It is known that the minimum rate of UAVs is maximized when the rates of UAVs are the same such as $R_1(\beta_1^o, \tau^o) = \cdots = R_K(\beta_K^o, \tau^o)$. 
This characteristic will be used for the design of the low-complexity iterative bandwidth-allocation algorithm.

In the $n$th iteration of the second phase, the BAF is re-allocated between the $\check{k}$th UAV and the $\hat{k}$th UAV to achieve   
\begin{align} \label{n_targ}
    \hat{R}(\boldsymbol{\beta}_n, \tau^o) = \check{R}(\boldsymbol{\beta}_n, \tau^o),
\end{align} 
where $\hat{k}$ is the index of the maximum rate UAV, and $\hat{R}(\boldsymbol{\beta}, \tau) = \max_{k}\left\{ R_{1}(\beta_1, \tau), \cdots, R_{K}(\beta_K, \tau) \right\}$ is the maximum rate of UAVs. 
To satisfy (\ref{n_targ}), the BAF is updated as
\begin{align}
\boldsymbol{\beta}_{n} = \boldsymbol{\beta}_{n-1} + \Delta \beta_n \mathbf{e}_n,
\end{align}
where 
\begin{align}
    \Delta \beta_n =  \beta_{\hat{k}} \left(\frac{\hat{R}(\boldsymbol{\beta}_{n-1}, \tau^o)- \check{R}(\boldsymbol{\beta}_{n-1}, \tau^o)}
    {2 \hat{R}(\boldsymbol{\beta}_{n-1}, \tau^o)} \right)
\end{align}
denotes the bandwidth allocation step size, and 
$\mathbf{e}_n = [0, \cdots, 1, \cdots, -1, \cdots, 0]$
denotes the index vector which has $1$ and $-1$ in the $\check{k}$th and the $\hat{k}$th indexes, respectively.  
The bandwidth-allocation algorithm will be performed repeatedly until 
$\mathcal{B}(\boldsymbol{\beta}_{n-1}, \tau^o)= \hat{R}(\boldsymbol{\beta}_{n-1}, \tau^o)- \check{R}(\boldsymbol{\beta}_{n-1}, \tau^o)  \leq \epsilon$ with a predefined threshold~$\epsilon$. 
The bandwidth allocation step size $\Delta \beta_n$ decreases as the number of iterations increases.
Therefore, we can state that the proposed algorithm always converges to the near-optimal TAF and BAF. The overall procedure of the proposed resource-allocation algorithm is described in Algorithm 1.

\begin{algorithm}[t]
\SetAlgoLined
\KwResult{ $\tau^o$, $\boldsymbol{\beta}^o$, $I_{\tau}$, $I_{\boldsymbol{\beta}}$ }
 {Initialize $\tau_\mathrm{l} = \epsilon$, $\tau_\mathrm{u} = 1-\epsilon$, $\boldsymbol{\beta}_0 = [1/K, \cdots, 1/K]$, and $n=0$}  \\
 \While{$ \tau_\mathrm{u} - \tau_\mathrm{l} > \epsilon$}{
  Update $n = n+1$ \\
  Compute the TAF $\tau_n = (\tau_\mathrm{l} + \tau_\mathrm{u})/2$ \\
  \eIf{ $ \frac{\partial \check{R}(\boldsymbol{\beta},\tau)}{\partial \tau}|_{\tau = \tau_n}  > 0$}{
   Update $\tau_\mathrm{l} = \tau_n$ 
   }{
   Update $\tau_\mathrm{u} = \tau_n$ 
  }
 }
 {Set $\tau^o = \tau_n$ and $I_{\tau} = n$ \\
 Update $n=1$ } \\
 \While{$ \mathcal{B}(\boldsymbol{\beta}_{n-1}, \tau^o) > \epsilon$}{
   {Compute the bandwidth allocation step size $\Delta \beta_n$  \\
   Generate the index vector $\mathbf{e}_n$  \\
   Update the BAF $\boldsymbol{\beta}_n = \boldsymbol{\beta}_{n-1} + \Delta\beta_n\mathbf{e}_n$  \\
   Update the rate of UAVs $R(\boldsymbol{\beta}_n, \tau^o)$ and $n = n+1$  }
 }
 {Set $\boldsymbol{\beta}^o = \boldsymbol{\beta}_{n-1}$ and $I_{\boldsymbol{\beta}} = n-1$}
   \caption{Proposed resource-allocation algorithm}
\end{algorithm}

The complexity of the proposed algorithm is represented as $\mathcal{O}(K+I_{\tau}+I_{\boldsymbol{\beta}}K)$ where 
$K$ is the number of UAVs,  
and $I_{\tau}$ and $I_{\boldsymbol{\beta}}$ are the required iterations for the optimal TAF and BAF, respectively.
Note that the complexity of using conventional bisection algorithm to solve (\ref{op1}) is $\mathcal{O}(I_{\tau}K+I_{\boldsymbol{\beta}}K I_{\beta_k})$ where $I_{\beta_k}$ is the required iterations for the optimal BAF of the $k$th UAV. This clearly shows that the complexity of proposed resource-allocation algorithm is always less than that of algorithm using the bisection method since $ I_{\beta_k} \gg 1$. 

\section{Simulation Results}
In this section, we present simulation results  
to confirm the closed-form outage probability of (\ref{Fana})
and to verify the advantage of the proposed resource-allocation algorithm. 
``Optimal'', ``Proposed'', ``Conventional'', and ``Equal-Bandwidth'' are presented for the performance comparison.
The ``Conventional'' algorithm computes the TAF and the BAF with the conventional bisection method \cite{Boyd}, 
and the ``Equal-Bandwidth'' algorithm sets the TAF to (\ref{t_s}) and the BAF to $\boldsymbol{\beta}=[1/K, \cdots, 1/K]$. 
We assume that the normalized duration of the RAP is $\nu_\mathrm{r}=0$ for ``Optimal'' to show the lower bound of the outage probability of the EH-UAV identification network.

The horizontal distances from the $k$th UAV and ground stations are set to be $d_{h,k}=\hat{d}\times k/K$ and $d_{g,k}=\hat{d} - d_{h,k}$ with the maximum horizontal distance $\hat{d}$. The altitude of the $k$th UAV is set to be $A_k = \hat{A}\times k/K$ for $k=1,...,K$ with the maximum altitude of UAVs $\hat{A}$. The simulation parameters are listed in Table. \ref{SP}.

\begin{table}
\caption{Simulation Parameters}
\label{SP}
\centering
\begin{tabular}{ c c |c c | c c }
\hline
 Parameter & Value & Parameter & Value & Parameter & Value \\
\hline
$f_\mathrm{c}$ [GHz] & 2.4 & $\sigma^2$ [dBm] & -114 & $B$ [MHz] & 1 \\
$N_\mathrm{s}$ & 10 & $N_{\mathrm{c}}$ & 4 & $N_{\mathrm{r}}$ & 4 \\
$\zeta$ & 0.7 & $\hat{d}$ [m] & 100 & $p_{\mathrm{c},k}$ [mW] & 100  \\
$m_{h,k}$ & 3 & $m_{g,k}$ & 3 & $a$ & 9.61 \\ 
$b$ & 0.16 & $\eta_{\text{LoS}}$ [dB]& 1 & $\eta_{\text{NLoS}}$ [dB] & 20 \\ $c$ [m/s] & $3 \times 10^8$  &  $\epsilon$ & $10^{-4}$  &  $R_{\mathrm{a}}$ [bps/Hz] & 1 \\
\hline
\end{tabular}
\end{table}

\begin{figure}
\centering
\includegraphics[width=5.4cm]{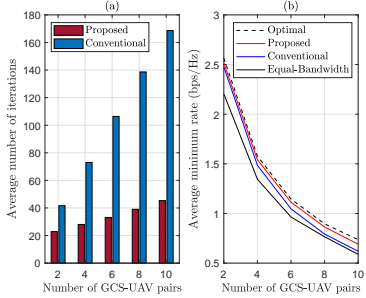} 
\caption{(a) Average number of iterations for convergence and (b) average minimum rate with the number of GCS-UAV pairs.}
\vspace{-11pt}
\label{fig4}
\end{figure}

Fig. 3(a) presents the average number of iterations for convergence,
and Fig. 3(b) shows the average minimum rate of UAVs as the number of GCS-UAV pairs. 
The maximum altitude of UAVs is $\hat{A}=120$ m, and the maximum velocity of UAVs is $\hat{V}=20$ m/s. 
It is shown that the proposed algorithm requires fewer iterations than the conventional algorithm
under the same threshold $\epsilon$. As the number of GCS-UAV pairs increases, the difference in the number of iterations between the proposed and the conventional algorithms increases. 
The proposed algorithm is less affected by the number of GCS-UAV pairs than the conventional algorithm since the proposed algorithm updates the rate of the UAVs for every iteration but the conventional algorithm updates the rate of the UAVs for every $K$ iterations.
It is also observed that the proposed algorithm outperforms both the conventional and the equal-bandwidth  algorithms in terms of the average minimum rate. 
Note that the resource-allocation algorithm which requires fewer iterations 
can provide higher rate of the UAVs due to the duration of the RAP.

Fig. 4 shows the outage probability with the maximum
altitude of UAVs when the number of GCS-UAVs pairs is
$K = 6$, and the required rate is $R_{\mathrm{a}}$ = 1 bps/Hz [13]. 
With the given deployment of UAVs, GCSs, and GRS, it is shown that the outage probabilities decrease with the maximum altitude of UAVs up to 90 m then decrease due to the A2G path-loss model in (\ref{pl}). 
It is observed that the closed-form expression of the outage probability derived in Section III matches exactly with the equal-bandwidth  algorithm.
It is also shown that the proposed algorithm not only provides near-optimal performance when the maximum velocity of UAVs is slow 
but also outperforms the conventional algorithm. 
Note that the optimal and the equal-bandwidth algorithms are irrelevant to the maximum velocity of UAVs because the normalized duration of the RAP is set to zero for these cases.
The performance gap between the proposed algorithm and the conventional algorithm 
becomes larger as the maximum velocity of UAVs increases because
the portion of the RAP is relatively large when the maximum velocity of UAVs is high.

\begin{figure}
\centering
\includegraphics[width=5.4cm]{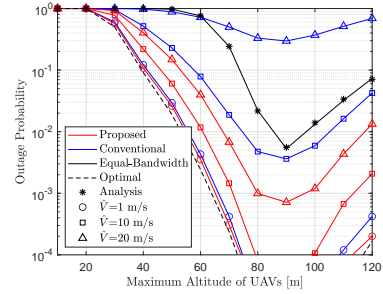} 
\caption{Outage probability as the maximum altitude of UAVs for various maximum velocities of UAVs.}
\vspace{-12pt}
\label{fig5}
\end{figure}

\section{Conclusion}
In this letter, we analyzed the outage probability to evaluate the performance of the EH-UAV identification network. 
We proposed a fast-converging algorithm for time and bandwidth allocation to optimize the outage probability.
Simulation results showed that the proposed algorithm provided near-optimal performance and required fewer iterations for convergence compared to the conventional algorithm. The performance gap between the proposed and the conventional algorithms became larger as the maximum velocity of UAVs increased due to the portion of the RAP in the block time.

\ifCLASSOPTIONcaptionsoff
  \newpage
\fi

\end{document}